\documentclass[lettersize,journal,12pt,onecolumn,draftclsnofoot]{IEEEtran}

\usepackage{cite}
\usepackage{textcomp}
\usepackage{xcolor}
\usepackage[normalem]{ulem} 

\usepackage{graphicx}
\ifCLASSOPTIONcompsoc 
    \usepackage[caption=false,font=normalsize,labelfont=sf,textfont=sf]{subfig}
\else
    \usepackage[caption=false,font=footnotesize]{subfig}
\fi

\usepackage{amsmath,amsfonts,amsthm}
\usepackage[T1]{fontenc}
\usepackage[cmintegrals]{newtxmath}
\usepackage{bm}

\usepackage{algorithm}
\usepackage{algorithmic}

\usepackage[acronym,,nohypertypes={acronym,notation}]{glossaries}

\newacronym{3gpp}{3GPP}{3rd Generation Partnership Project}
\newacronym{5gnr}{5G NR}{5G new radio}
\newacronym{awgn}{AWGN}{additive white Gaussian noise}
\newacronym{bs}{BS}{base station}
\newacronym{cdf}{CDF}{cumulative distribution function}
\newacronym{ci}{CI}{cancellation indication}
\newacronym{csi}{CSI}{channel state information}
\newacronym{dl}{DL}{downlink}
\newacronym{embb}{eMBB}{enhanced mobile broadband}
\newacronym{5g}{5G}{fifth generation}
\newacronym{iid}{i.i.d.}{independent and identically distributed}
\newacronym{in}{IN}{interference nulling}
\newacronym{kpi}{KPI}{key performance indicator}
\newacronym{ls}{LS}{least squares}
\newacronym{lhs}{LHS}{left-hand side}
\newacronym{los}{LoS}{line-of-sight}
\newacronym{mimo}{MIMO}{multiple-input multiple-output}
\newacronym{mmimo}{M-MIMO}{massive MIMO}
\newacronym{nr}{NR}{new radio}
\newacronym{nlos}{NLoS}{non-line-of-sight}
\newacronym{noma}{NOMA}{non-orthogonal multiple access}
\newacronym{ofdm}{OFDM}{orthogonal frequency-division multiplexing}
\newacronym{oma}{OMA}{orthogonal multiple access}
\newacronym{qos}{QoS}{quality of service}
\newacronym{rb}{RB}{resource block}
\newacronym{rf}{RF}{radio frequency}
\newacronym{rhs}{RHS}{right-hand side}
\newacronym{rx}{Rx}{receiver}
\newacronym[plural=RISs]{ris}{RIS}{reconfigurable intelligent surface}
\newacronym{risa}{RISA}{RIS' array}
\newacronym{risc}{RISC}{RIS' controller}
\newacronym{risru}{RISRu}{RIS' radio unit}
\newacronym{se}{SE}{spectral efficiency}
\newacronym{sic}{SIC}{successive interference cancellation}
\newacronym{sinr}{SINR}{signal-to-interference-plus-noise ratio}
\newacronym{snr}{SNR}{signal-to-noise ratio}
\newacronym{sr}{SR}{scheduling request}
\newacronym{tti}{TTI}{transmission time interval}
\newacronym{tx}{Tx}{transmitter}
\newacronym[plural=UEs, firstplural=users' equipment (UEs)]{ue}{UE}{user's equipment}
\newacronym{pr}{PR}{phasors rotation}
\newacronym{ul}{UL}{uplink}
\newacronym{upa}{UPA}{uniform planar array}
\newacronym{urllc}{URLLC}{ultra-reliable-low-latency communication}

\newcommand{\complexset}{\mathbb{C}}

\newcommand{\integerset}{\mathbb{Z}}

\begin{document}

\title{
    Uplink Multiplexing of eMBB{\slash}URLLC Services Assisted by Reconfigurable Intelligent Surfaces
}

\author{
    João Henrique Inacio de Souza,
    Victor Croisfelt,
    Rados{\l}aw Kotaba,
    Taufik Abrão,
    and Petar Popovski

    \thanks{
    		This work has been submitted to the IEEE for possible publication. Copyright may be transferred without notice, after which this version may no longer be accessible.
    		
        J. H. Inacio de Souza and T. Abrão are with the Department of Electrical Engineering, Universidade Estadual de Londrina, Brazil (e-mail: joaohis@outlook.com, taufik@uel.br).

        V. Croisfelt, R. Kotaba, and P. Popovski are with the Department of Electronic Systems, Aalborg University, Denmark (e-mail: \{vcr,rak,petarp\}@es.aau.dk).
    }
}

\maketitle

\begin{abstract}
    This letter proposes a scheme assisted by a reconfigurable intelligent surface (RIS) for efficient uplink traffic multiplexing between  enhanced mobile broadband (eMBB) and ultra-reliable-low-latency communication (URLLC). The scheme determines two RIS configurations based only on the eMBB channel state information (CSI) available at the base station (BS). The first optimizes eMBB quality of service, while the second reduces eMBB interference in URLLC traffic by temporarily silencing the eMBB traffic. Numerical results demonstrate that this approach, relying solely on eMBB CSI and without BS coordination, can outperform the state-of-the-art preemptive puncturing by 4.9 times in terms of URLLC outage probability.
\end{abstract}

\begin{IEEEkeywords}
    Reconfigurable intelligent surface (RIS), enhanced mobile broadband (eMBB), ultra-reliable low-latency communications (URLLC), and multiplexing.
\end{IEEEkeywords}

\section{Introduction}\label{sec:introduction}

\IEEEPARstart{T}{he} \gls{5g} of mobile networks have been launched with features to support heterogeneous services with different \gls{qos} requirements and traffic characteristics~\cite{Chowdhury2020}. In particular, \gls{embb} services require extremely high \gls{se}, while \gls{urllc} services demand high reliability and low latency~\cite{Popovski2018}. Due to these very different \glspl{qos}, there is a need for new multiplexing strategies that perform effectively in challenging channel conditions and better accommodate heterogeneous traffic demands from both \gls{embb} and \gls{urllc} services.

An \gls{ris} is a thin sheet of composite material that can cover, \emph{e.g.}, parts of walls and buildings. It can reflect incident signals to desired directions by dynamically configuring the phase shifts of the many elements that compose it~\cite{DiRenzo2020}. In the literature, works like~\cite{Xie2021,Li2023,Ren2022} study the fundamental limits and optimize the \gls{qos} in \gls{ris}-assisted \gls{urllc} applications. 
Our work is motivated by the fact that \gls{ris}
can help to deliver the \gls{qos} for heterogeneous services, an aspect that has received less attention in the research literature.

Regarding heterogeneous traffic multiplexing, \gls{5gnr} has introduced the \emph{preemptive puncturing scheme}, that relies on \gls{bs} coordination to interrupt the \gls{embb} traffic and prioritizes the \gls{urllc} transmissions~\cite{3gpp38300,Almekhlafi2022-mar}. Despite its fair \gls{dl} performance, in the \gls{ul}, such a strategy increases the \gls{urllc} latency, mainly due to the waiting time for the \gls{bs} to grant scheduling responses. Meanwhile, in the case of \gls{ris}-assisted systems, the multiplexing of \gls{embb}\slash\gls{urllc} services gives rise to two additional challenges~\cite{Almekhlafi2022-mar}. First, the \gls{bs} is unaware of the time of arrival of a \gls{urllc} packet and cannot estimate its \gls{csi} to tailor the \gls{ris} configuration~\cite{Popovski2018}. Second, controlling the \gls{ris} adds extra overhead that can violate the \gls{urllc} latency requirements.

For the \gls{dl}, some of the deployment issues associated with \gls{ris} can be mitigated through the \gls{ris} configurations and resource allocation policies proposed in~\cite{Zarini2023, Almekhlafi2022-feb, Ghanem2022}. In contrast, for the \gls{ul}, \cite{Souto2021} suggests \gls{ris} configuration designs to aid both \gls{embb} and \gls{urllc} \gls{ue} even when \gls{urllc} \gls{csi} is unavailable, using \gls{noma} for service multiplexing. However, these studies still rely on heterogeneous multiplexing strategies that require \gls{bs} coordination and may face challenges related to \gls{ris} control, without directly addressing the previously mentioned issues.

In this letter, we propose a new \gls{ris}-assisted multiplexing scheme to support heterogeneous \gls{embb} and \gls{urllc} \gls{ul} traffic.  Rather than depending on \gls{bs} coordination, our scheme is based on the assumption that the \gls{ris} is equipped with an antenna, and the \gls{ris} is then capable of minimally processing the signal received by this antenna to detect \gls{urllc} traffic. Then, inspired by the preemptive puncturing of \gls{5gnr}, the \gls{ris} multiplexes the services using two configurations, which rely solely on \gls{embb} \gls{csi}. The first assists the \gls{embb} \gls{ue} by optimizing the signal strength and thus its \gls{se}.
The second, motivated by~\cite{Jiang2022}, is designed to assist the \gls{urllc} \gls{ue} (if detected) by temporarily silencing the \gls{embb} interference. The proposed scheme is compared to benchmarks in terms of outage probability for the \gls{urllc} service and \gls{se} for the \gls{embb}. {Numerical results show that the proposed scheme can outperform the outage probability of preemptive puncturing by 4.9~times.}

\section{System Model}\label{sec:system-model}

We consider a narrowband \gls{ul} channel of a wireless system with one single-antenna \gls{bs}, one hybrid \gls{ris}, and two single-antenna \glspl{ue}, where the \gls{bs} controls the \gls{ris} via an out-of-band control channel~\cite{Bjornson2022}. One of the \gls{ue} uses the \gls{embb} service, while the other uses the \gls{urllc} one.\footnote{To attain efficient multi-user communication, it is common to multiplex several \glspl{ue} of each traffic type in a wideband channel. However, to simplify the analysis, we consider a single \gls{ue} of each traffic type in a narrowband channel, leaving the general case for a future extension of this work.} Henceforth, we index the \glspl{ue} by $\iota \in \{\text{e},\text{u}\}$, where $\text{e}$ and $\text{u}$ refer to \gls{embb} and \gls{urllc}, respectively. We assume an industrial scenario where the \gls{los} between the \gls{bs} and the \glspl{ue} is blocked by obstacles. Thus, the \gls{ris} is deployed to simultaneously have \gls{los} to the \gls{bs} and the \glspl{ue}. The \gls{ris} is a square \gls{upa} of $N \in \integerset_+$ half-wavelength spaced passive reflecting elements and one active element, placed at the center of the \gls{ris}. The \gls{ris} can perform light computational tasks by processing the signal received by its active element. Each passive element can induce a phase shift $\theta_n \in [0, 2\pi)$ to an impinging signal with marginal impact on its amplitude. On this basis, we define an \gls{ris} configuration as $\boldsymbol{\psi} = [\psi_1 ~ \cdots ~ \psi_N]^\transp$, where $\psi_n = e^{-j \theta_n}$ is the reflection coefficient of the $n$-th \gls{ris} element.

\subsection{Structure of a Uplink Frame}\label{sec:structure-uplink-frame}

\begin{figure}[h]
\centering
\includegraphics[width=.5\columnwidth]{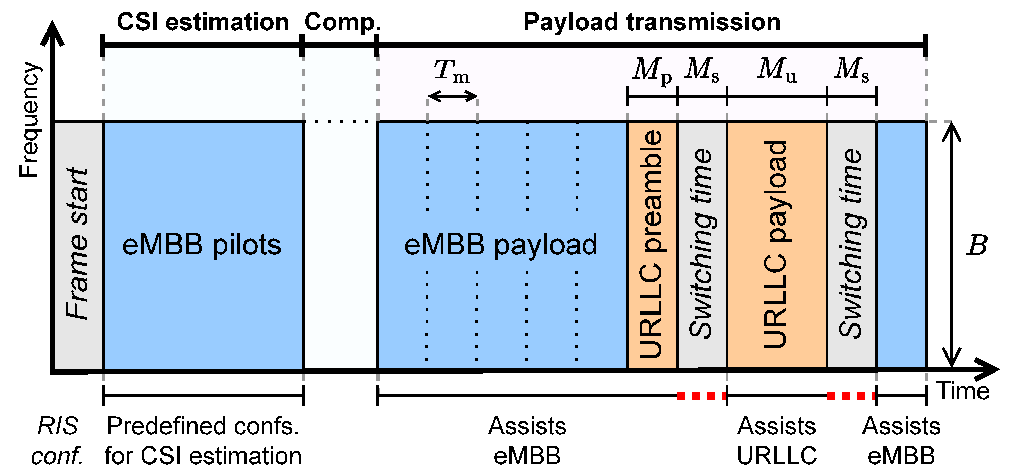}
\vspace{-7mm}
\caption{\Gls{ul} frame divided into \gls{csi} estimation, computing, and payload transmission.}
\label{fig:ul-frame}
\end{figure}

Fig.~\ref{fig:ul-frame} depicts a \gls{ul} frame, in which the time-frequency resources used by the \glspl{ue} are organized into a grid. In the time domain, the length of the \gls{ul} frame is $T > 0$, and it is divided into $M \in \integerset_+$ identical mini-slots of duration $T_\text{m} > 0$, which are indexed by $\mathcal{M}=\{1, \dots, M\}$. In the frequency domain, the spectrum reserved for \gls{ul} has the bandwidth $B > 0$.  Due to different requirements, the \glspl{ue} have different \glspl{tti}. To achieve high \gls{se}, the \gls{embb} \gls{tti} spans over the entire frame. Conversely, the \gls{urllc} \gls{tti} spans over a limited number of $M_\text{u} \in \integerset_+, ~ 1 \leq M_\text{u} \leq M$ contiguous mini-slots to guarantee the low-latency requirement. We denote as $\mathcal{M}_\text{u} \subseteq \mathcal{M}$ with $|\mathcal{M}_\text{u}| = M_\text{u}$ the set of mini-slots of the \gls{urllc} \gls{tti}. Moreover, we consider that the \gls{embb} \gls{ue} is admitted and scheduled by the \gls{bs} at the frame start, while the \gls{urllc} \gls{ue} can start transmitting at any mini-slot.

\subsection{Signal Model}

Let the vector $\mathbf{h}_\text{b} \in \complexset^N$ denotes the \gls{bs}-\gls{ris} channels and the vectors $\mathbf{h}_\iota \in \complexset^N$ denote the \gls{ris}-\glspl{ue} channels. {We further assume the block-fading model, \emph{i.e.}, the channels remain constant for the entire frame duration.} Therefore, the \gls{bs}-\glspl{ue} cascaded channels are represented by the vector~\cite{Jiang2022}
\begin{equation}
    \label{eq:ues-ris-channels}
    \mathbf{g}_\iota \triangleq \mathrm{diag} (\mathbf{h}_\text{b}) \, \mathbf{h}_\iota.
\end{equation}
Finally, we let the scalars $g_{\iota,n} = [\mathbf{g}_\iota]_n$ denote the \gls{bs}-\glspl{ue} cascaded channels that pass through the $n$-th \gls{ris} element. We denote as $\mathbf{x}_{\iota,m} \in \complexset^{L}$ the vectors with the $L \in \integerset_+$ symbols transmitted by each \gls{ue} such that $\mathbb{E} \{ \| \mathbf{x}_{\iota,m} \|_2^2 \} = 1$ in the $m$-th mini-slot. Thus, $\mathbf{x}_{\text{u},m} = \mathbf{0}$ in the mini-slots with absence of \gls{urllc} traffic, that is, when $m \notin \mathcal{M}_\text{u}$. Let $p_\iota > 0$ denote the \glspl{ue}' transmit power for one mini-slot. Considering the \gls{ris} configuration $\boldsymbol{\psi}$, the received signal $\mathbf{y}_m (\boldsymbol{\psi}) \in \complexset^L$ at the \gls{bs} can be written as:
\begin{equation}
    \label{eq:received-signal}
    \mathbf{y}_m (\boldsymbol{\psi}) \triangleq \sqrt{p_\text{u}} ( \mathbf{g}_\text{u}^\htransp \boldsymbol{\psi} ) \mathbf{x}_{\text{u},m} + \sqrt{p_\text{e}} ( \mathbf{g}_\text{e}^\htransp \boldsymbol{\psi} ) \mathbf{x}_{\text{e},m} + \mathbf{w}_m,
\end{equation}
where $\mathbf{w}_m \sim \mathcal{CN} (\mathbf{0}, \sigma^2 \mathbf{I})$ is the \gls{awgn}.
Using~\eqref{eq:received-signal}, the \gls{snr} for the \gls{embb} \gls{ue} in the mini-slots without \gls{urllc} traffic is:
\begin{equation}
    \label{eq:embb-snr}
    \Gamma_{\text{e},m} (\boldsymbol{\psi}) \triangleq p_\text{e} | \mathbf{g}_\text{e}^\htransp \boldsymbol{\psi} |^2 / \sigma^2, ~ \forall m \in \mathcal{M} \backslash \mathcal{M}_\text{u}.
\end{equation}
Also, during the \gls{urllc} \gls{tti}, the \gls{sinr} for the \gls{urllc} \gls{ue} is given by
\begin{equation}
    \label{eq:urllc-sinr}
    \Gamma_{\text{u},m} (\boldsymbol{\psi}) \triangleq p_\text{u} | \mathbf{g}_\text{u}^\htransp \boldsymbol{\psi}|^2 / (p_\text{e} | \mathbf{g}_\text{e}^\htransp \boldsymbol{\psi} |^2 + \sigma^2), ~ \forall m \in \mathcal{M}_\text{u}.
\end{equation}
With~\eqref{eq:embb-snr} and~\eqref{eq:urllc-sinr}, one can measure the quality of the radio links established for the \glspl{ue} by the \gls{ris}.

\section{RIS-Assisted Multiplexing Scheme}\label{sec:scheme}

Here, we introduce our proposed multiplexing scheme.
First, we explain the three phases comprising a \gls{ul} frame and give an overview of how our scheme works. Then, we discuss how the \gls{ris} could detect the \gls{urllc} traffic, and we parameterize the performance of an arbitrary detector by its miss detection rate, $\epsilon_\text{m}$. Finally, to multiplex the \gls{embb} and \gls{urllc} services, we design two \gls{ris} configurations that rely only on \gls{embb} \gls{csi}.

\subsection{Phases of the Multiplexing Scheme}

From Fig.~\ref{fig:ul-frame}, a \gls{ul} frame is divided into three phases.
\textbf{1.~\Gls{csi} Estimation:} The \gls{bs} estimates the \gls{embb} \gls{csi} by using pilot symbols sent by the \gls{embb} \gls{ue}, while the \gls{ris} switches among predefined configurations (for more details, see ~\cite{DiRenzo2020}).
\textbf{2.~Computing:} The \gls{bs} computes two \gls{ris} configurations based on the \gls{embb} \gls{csi}, the \emph{\gls{embb}-oriented configuration} and the \emph{\gls{urllc}-oriented configuration} to potentially multiplex the \gls{embb} and \gls{urllc} services within the given frame. Finally, these two configurations are sent to the \gls{ris} to be stored there for potential use in the next phase.
\textbf{3.~Payload Transmission:} In this phase, the scheduled \gls{embb} \gls{ue} transmits its payload whereas the \gls{ris} starts by being configured with the \gls{embb}-oriented configuration. At any time in this phase, if the \gls{urllc} \gls{ue} has a packet to send, it will start its \gls{tti} by sending a preamble over $M_\text{p} \in \integerset_+$ mini-slots to announce its intention to transmit payload data. By exploiting its active element\footnote{In practice, this active element could also be the antenna of the communication interface used to control the \gls{ris}~\cite{DiRenzo2020}, making it a viable solution.} and the \gls{urllc} preamble, the \gls{ris} can \textit{detect} the start of the \gls{urllc} traffic. If \gls{urllc} traffic is detected, the \gls{ris} switches to the \gls{urllc}-oriented configuration to multiplex the \gls{urllc} \gls{tti}; otherwise, it keeps the \gls{embb}-oriented configuration. In the former case, when the \gls{urllc} \gls{tti} is over, the \gls{ris} switches back to the \gls{embb}-oriented configuration. Note that while all this occurs, the \gls{embb} \gls{tti} is not interrupted, so part of the \gls{embb} mini-slots will be erased by the \gls{urllc} traffic, and by the changes in the environment caused by the \gls{ris}.\footnote{To decode the \gls{embb} payload data at the \gls{bs} with a limited number of erased mini-slots, erasure coding can be applied by the \gls{embb} \gls{ue}~\cite{Popovski2018}.}

In practice, the \gls{ris} can impose a physical latency due to switching configurations. By assuming that the time the \gls{ris} takes to change its phase shifts is around some microseconds~\cite{DiRenzo2020}, we consider the \gls{ris} takes $M_\text{s} \in \integerset_+$ mini-slots to switch between configurations. We also assume that the \gls{urllc} \gls{ue} is aware of this \emph{switching time} and waits for it before transmitting the payload. Hence, in case the \gls{urllc} traffic is detected, the \gls{ris} adds an overhead of $M_{\text{s}}$ mini-slots for the \gls{urllc} \gls{ue} and $2M_{\text{s}}$ mini-slots for the \gls{embb} one.

\subsection{Detecting the URLLC Traffic}

The \gls{urllc} traffic can be identified at the \gls{ris} using the preamble signaling spanning over $M_{\rm p}$ mini-slots. To do so, one effective approach would involve the joint design of the preamble and detection algorithm to meet \gls{urllc} \glspl{qos}. For instance, in the context of \gls{urllc} service, a robust scheme presented in~\cite{Fischione2019} introduces a brief preamble consisting of a single \gls{ofdm} symbol, detected at the receiver using differential detection. This algorithm can be seamlessly implemented in a hybrid \gls{ris} with limited signal processing capacity. However, we investigate the performance of an arbitrary detection scheme, assessing its general impact on multiplexing efficiency without relying on a specific scheme. We parameterize the performance of this detector by its \emph{miss detection rate}, defined as $0 \leq \epsilon_\text{m} \leq 1$, whose impact is evaluated in Section \ref{sec:numerical-results}.

\subsection{Computing the eMBB-Oriented RIS Configuration}\label{sec:scheme:embb}

The \gls{embb}-oriented \gls{ris} configuration $\boldsymbol{\psi}^\text{e} \in \complexset^N$ is set to maximize the \gls{embb} \gls{snr}, which can be posed as the following optimization problem
\begin{equation}
    \label{eq:coherent-passive-beamforming}
    \underset{\boldsymbol{\psi} \in \complexset^N}{\max} \; | \mathbf{g}_\text{e}^\htransp \boldsymbol{\psi} |^2 / \sigma^2, \; \mathrm{s.t.} \; \mathrm{C}_1: | \psi_n | = 1, ~ \forall n \in \{1, \dots, N\}.
\end{equation}
Fortunately, this problem assumes a family of closed-form solutions known as coherent passive beamformers~\cite{Bjornson2022}. Based on this and with \gls{embb} \gls{csi},\footnote{We assume perfect \gls{csi} knowledge at the \gls{bs} to show the upper bound performance of the proposed scheme, leaving the study of the impact of imperfect \gls{csi} for future work.} $\boldsymbol{\psi}^\text{e}$ can be computed as:
\begin{gather}
    \label{eq:ris-configuration-embb}
    \boldsymbol{\psi}^\text{e} = [ e^{-j \theta_1^\text{e}} ~ \cdots ~ e^{-j \theta_N^\text{e}} ]^\transp, \text{ where}\\
    \nonumber
    \theta_n^\text{e} = - \mathrm{arg}(g_{\text{e},n})+ \bar{\theta},~n \in \{1, \dots, N\},
\end{gather}
and $\bar{\theta} \in [0, 2\pi)$ due to the phase periodicity. For the sake of simplicity but without compromising generality, we take $\bar{\theta} = 0$. In such a case, from~\eqref{eq:ues-ris-channels}, the effective channel of the \gls{embb} \gls{ue} can be simplified to $\mathbf{g}_\text{e}^\htransp \boldsymbol{\psi}^\text{e} = \sum_{n = 1}^N |g_{\text{e},n}|$. Hence, the \gls{ris} yields the maximum array gain of $N^2$, providing high \gls{se}.

\subsection{Computing the URLLC-Oriented RIS Configurations}\label{sec:scheme:urllc}

At the \gls{bs}, the computation of an \gls{ris} configuration to multiplex the \gls{urllc} \gls{tti} is challenging since only \gls{embb} \gls{csi} is available at the \gls{bs}. Our idea is to find a configuration that mitigates the interference caused by the \gls{embb} traffic to the \gls{urllc} one, that is, that temporarily silences the \gls{embb} \gls{ue}. We present two methods for computing the \gls{urllc}-oriented configuration. The first one is a heuristic based on \emph{phasors rotation} that tries to cancel out the channel gain of the \gls{embb} \gls{ue} by compensating the phase shifts of the \gls{ris} elements via subtraction. The second one is an \emph{alternating projection} algorithm to approximate a configuration that nulls the \gls{embb} interference.

\vspace{1mm}
\noindent\textbf{Method 1: \textit{\Gls{pr}}}.
Given the \gls{embb} \gls{csi}, let us define the problem of finding an \gls{ris} configuration that minimizes the channel gain of the \gls{embb} \gls{ue} as
\begin{equation}
    \label{eq:minimize-embb-effective-channel-gain}
    \underset{\boldsymbol{\psi} \in \complexset^N}{\min} \; | \mathbf{g}_\text{e}^\htransp \boldsymbol{\psi} |, \; \mathrm{s.t.} \; \mathrm{C}_1,
\end{equation}
with $\mathrm{C}_1$ as in \eqref{eq:coherent-passive-beamforming}, which makes the problem to be not convex.
However, notice that the channel gain is lower-bounded such that $| \mathbf{g}_\text{e}^\htransp \boldsymbol{\psi} | \geq 0$. Therefore, our goal is to find a configuration that make this gain as close as possible to zero.

We start by presenting a heuristic algorithm with low computational complexity, which yields a sub-optimal solution to problem~\eqref{eq:minimize-embb-effective-channel-gain}. The heuristic is based on the representation of the \gls{embb} cascaded channels as phasors, and the idea that we can individually rotate them so that they cancel each other out, nulling the \gls{embb} channel. This is made by dividing the \gls{ris} elements into two sets, where the goal of one of the sets is to eliminate the contribution of the other. 
From~\eqref{eq:ues-ris-channels}, let the $n$-th \gls{embb} cascaded channel reflected by the \gls{ris} be represented by the phasor such that
\begin{equation}
    \label{eq:phasor}
    A_n e^{j \omega_n} \triangleq g_{\text{e},n}^* \psi_n,
\end{equation}
where $A_n = |g_{\text{e},n}|$ is the phasor's amplitude, and $\omega_n = - \mathrm{arg} (g_{\text{e},n}) - \theta_n$ is the phasor's angle.
Let the nonempty sets $\mathcal{N}_0$ and $\mathcal{N}_\pi$ denote a partition of the \gls{ris} elements, \textit{i.e.} $\mathcal{N}_0 \cup \mathcal{N}_\pi = \{1, \dots, N\}$ and $\mathcal{N}_0 \cap \mathcal{N}_\pi = \emptyset$.
The \gls{ris} elements are configured with the following phase shift according to the set they belong
\begin{equation}
    \label{eq:phase-shifts-phasor-rotation}
        \theta_n^\text{u} = \begin{cases}
        - \mathrm{arg}(g_{\text{e},n}), & n \in \mathcal{N}_0\\
        \pi - \mathrm{arg}(g_{\text{e},n}), & n \in \mathcal{N}_\pi
    \end{cases}.
\end{equation}
Using the configuration $\boldsymbol{\psi}^\text{u} = [e^{-j \theta_1^\text{u}} ~ \cdots ~ e^{-j \theta_N^\text{u}}]^\transp$, the cascaded channels belonging to sets $\mathcal{N}_0$ and $\mathcal{N}_\pi$ are out of phase, since $\omega_n = 0$ if $n \in \mathcal{N}_0$ and $\omega_n = -\pi$ if $n \in \mathcal{N}_\pi$. Therefore, from~\eqref{eq:phasor} and~\eqref{eq:phase-shifts-phasor-rotation}, the \gls{embb} channel gain with $\boldsymbol{\psi}^\text{u}$ is equal to
\begin{equation}
    \label{eq:effective-channel-embb}
    \textstyle
    | \mathbf{g}_\text{e}^\htransp \boldsymbol{\psi}^\text{u} | = | \sum_{n = 1}^N A_n e^{j \omega_n} | = | \sum_{n \in \mathcal{N}_0} A_n - \sum_{n' \in \mathcal{N}_\pi} A_{n'} |.
\end{equation}
Thus, to mitigate the interference caused by the \gls{embb} traffic, we need to determine the sets $\mathcal{N}_0$ and $\mathcal{N}_\pi$ that approximately null~\eqref{eq:effective-channel-embb}.
Since each cascaded channel can belong to either $\mathcal{N}_0$ or $\mathcal{N}_\pi$, minimizing~\eqref{eq:effective-channel-embb} is a combinatorial optimization problem with $2^N$ candidate solutions.
As this is not tractable even for moderate size \glspl{ris}, in Algorithm~\ref{alg:minimize-effective-channel} we present an intuitive method for determining $\mathcal{N}_0$ and $\mathcal{N}_\pi$, approximating a solution that minimizes~\eqref{eq:effective-channel-embb} in feasible computation time.

\begin{algorithm}[h]
    \caption{Phasors rotation algorithm.}
    \label{alg:minimize-effective-channel}
        \begin{algorithmic}[1]
            \renewcommand{\algorithmicrequire}{\textbf{input:}}
            \renewcommand{\algorithmicensure}{\textbf{output:}}  
            \REQUIRE The channel vector $\mathbf{g}_\text{e}$
            \ENSURE The \gls{ris} configuration $\boldsymbol{\psi}^\text{u}$
            \STATE $A_n \gets |g_{\text{e},n}|$
            \STATE $(\alpha_i)_{i = 1}^N \gets \mathrm{sort} ( A_1, \dots, A_N )$
            \STATE $N^* \gets \mathrm{arg \; min} \; \{ | \sum_{i = 1}^{N'} \alpha_i - \sum_{i' = N' + 1}^{N} \alpha_{i'} | \mid 1 \leq N' \leq N \}$
            \STATE $\mathcal{N}_0 \gets \{\mu(1), \dots, \mu(N^*)\}$, $\mathcal{N}_\pi \gets \{\mu(N^* + 1), \dots, \mu(N)\}$
            \STATE Set $\theta_n^\text{u}$, $\forall n \in \{1,\dots,N\}$ according to eq.~\eqref{eq:phase-shifts-phasor-rotation}
            \RETURN $\boldsymbol{\psi}^\text{u} \gets [e^{-\theta_1^\text{u}} ~ \cdots ~ e^{-\theta_N^\text{u}}]^\transp$
        \end{algorithmic}
\end{algorithm}

The algorithm works as follows. Initially, the amplitudes of the phasors representing the cascaded channels are computed. Then, the algorithm finds an integer number $1 \leq N^* \leq N$ such that the sum of the amplitudes of the $N^*$ shortest phasors is as close as possible to the sum of the amplitudes of the $N - N^*$ remaining ones.
Finally, the set $\mathcal{N}_0$ is created with the indices of the elements associated with the $N^*$ shortest phasors, while $\mathcal{N}_\pi$ is created with the indices of the remaining ones.
In the algorithm, $\mu(\cdot)$ maps the indices of $(\alpha_i)_{i = 1}^N$ to $\{A_n\}_{n = 1}^N$.

\vspace{1mm}
\noindent\textbf{Method 2: \textit{Interference Nulling (IN)}}.
The second method of finding an \gls{ris} configuration that nulls the \gls{embb} interference at the \gls{bs} can be cast as the following feasibility problem
\begin{equation}
    \label{eq:interference-nulling}
    \mathrm{find} \; \boldsymbol{\psi} \in \complexset^N, \; \mathrm{s.t.} \; \mathrm{C}_1 \text{ and } \mathrm{C}_2 : \mathbf{g}_\text{e}^\htransp \boldsymbol{\psi} = 0,
\end{equation}
where $\mathrm{C}_1$ is defined as in~\eqref{eq:coherent-passive-beamforming}, while $\mathrm{C}_2$ ensures that the \gls{embb} traffic does not interfere at the \gls{bs}. Similarly to~\eqref{eq:minimize-embb-effective-channel-gain}, this problem is not convex due to the unit modulus constraints~$\mathrm{C}_1$.
However, notice that, if any, a solution to the problem is any vector that belongs to the intersection between the sets defined by constraints~$\mathrm{C}_1$ and~$\mathrm{C}_2$.
In this case, the alternating projection algorithm can approximate a solution with a high probability of convergence.
In this sense, Algorithm~\ref{alg:interference-nulling} presents an adaption of the alternating projection algorithm in~\cite{Jiang2022} that iteratively approximates a solution to problem~\eqref{eq:interference-nulling}.
During each iteration $t \in \integerset_+$, the vector of reflection coefficients $\boldsymbol{\psi}_{t-1}$ is projected sequentially onto the set $\{ \boldsymbol{\psi} \in \complexset^N \mid \mathbf{g}_\text{e}^\htransp \boldsymbol{\psi} = 0 \}$, then onto the set $\{ \boldsymbol{\psi} \in \complexset^N \mid |\psi_n| = 1, \forall n \in \{1,\dots,N\} \}$. The projection operators consider the smallest Euclidean distance from $\boldsymbol{\psi}_{t-1}$ to a point in the projected set, derived according to \cite[eq. (20)]{Jiang2022}.
We use early stopping and maximum iterations as stopping criteria.

\begin{algorithm}[h]
    \caption{Alternating projection algorithm for IN.}
    \label{alg:interference-nulling}
    \begin{algorithmic}[1]
        \renewcommand{\algorithmicrequire}{\textbf{input:}}
        \renewcommand{\algorithmicensure}{\textbf{output:}}
        \REQUIRE The channel vector $\mathbf{g}_\text{e}$ and the initial configuration $\boldsymbol{\psi}_0$
        \ENSURE The \gls{ris} configuration $\boldsymbol{\psi}^\text{u}$
        \STATE $\mathbf{v} \gets \mathbf{g}_\text{e} / \| \mathbf{g}_\text{e} \|_2^2$, $t \gets 1$
        \REPEAT
        \STATE $\tilde{\boldsymbol{\psi}} \gets \boldsymbol{\psi}_{t-1} - ( \mathbf{v}^\htransp \boldsymbol{\psi}_{t-1} ) \mathbf{v}$
        \STATE $\boldsymbol{\psi}_t \gets [ \tilde{\psi}_1/|\tilde{\psi}_1| ~ \cdots ~ \tilde{\psi}_N/|\tilde{\psi}_N| ]^\transp$
        \STATE $t \gets t + 1$
        \UNTIL{stopping criterion is satisfied}
        \RETURN $\boldsymbol{\psi}^\text{u} \gets \boldsymbol{\psi}_{t-1}$
    \end{algorithmic}
\end{algorithm}

\section{Analysis}

This section introduces the metrics used to analyze the proposed multiplexing scheme.


\vspace{1mm}
\noindent\textbf{Performance Analysis:} By using the \gls{ris} configurations of Section \ref{sec:scheme}, we present expressions for the outage probabilities achieved by the \gls{embb} and \gls{urllc} \glspl{ue}. Initially, considering the configuration $\boldsymbol{\psi}^\text{e}$ in~\eqref{eq:ris-configuration-embb}, the instantaneous mutual information per mini-slot of the \gls{embb} data stream at the \gls{bs} is:
\begin{align}
    \nonumber
    I_\text{e} (p_\text{e}) & \triangleq M^{-1}
    \textstyle \sum_{m = 1}^M ( 1 - \xi ) \log_2 (1 + \Gamma_{\text{e},m} (\boldsymbol{\psi}^{\text{e}})),\\
    \label{eq:mutual-information-embb}
    & \textstyle = ( 1 - \xi ) 
    \log_2 ( 1 + p_\text{e} ( \sum_{n = 1}^N |g_{\text{e},n}| )^2 / \sigma^2 ),
\end{align}
where the pre-log term $\xi = (M_\text{p} + 2 M_\text{s} + M_\text{u}) / M$ accounts for the mini-slots of the \gls{urllc} \gls{tti} and \gls{ris} configuration switching.
Similarly, the instantaneous mutual information per mini-slot of the \gls{urllc} data stream considering the \gls{ris} configuration $\boldsymbol{\psi}^\text{u}$ computed by Algorithms~\ref{alg:minimize-effective-channel} or~\ref{alg:interference-nulling} is derived as
\begin{align}
    \nonumber
    I_\text{u} (p_\text{u}, p_\text{e}) & \triangleq M^{-1}_\text{u} \textstyle \sum_{m \in \mathcal{M}_\text{u}} \log_2 (1 + \Gamma_{\text{u},m} (\boldsymbol{\psi}^{\text{u}})),\\
    & = \log_2 ( 1 + p_\text{u} | \mathbf{g}_\text{u}^\htransp \boldsymbol{\psi}^\text{u} |^2 / (p_\text{e} | \mathbf{g}_\text{e}^\htransp \boldsymbol{\psi}^\text{u} |^2 + \sigma^2) ).
\end{align}
Therefore, given constant transmit power, the outage probabilities of the \gls{embb} and \gls{urllc} \glspl{ue} as functions of the \glspl{se} per mini-slot, $r_\iota > 0$, are respectively equal to
\begin{gather}
    \nonumber
    P_\text{e} (r_\text{e}) \triangleq \mathrm{Pr}(I_\text{e} (p_\text{e}) < r_\text{e}) \text{ and }
    P_\text{u} (r_\text{u}) \triangleq \mathrm{Pr}(I_\text{u} (p_\text{u}, p_\text{e}) < r_\text{u}).
\end{gather}


\vspace{1mm}
\noindent\textbf{Latency Analysis:}
From Fig.~\ref{fig:ul-frame}, the latency introduced by the proposed scheme to the \gls{urllc} traffic is governed by: \textbf{a)}~the transmission of the \gls{urllc} preamble, \textbf{b)}~the delay due to processing of the preamble and switching between configurations at the \gls{ris}, and \textbf{c)}~the transmission of the \gls{urllc} payload symbols. Neglecting the propagation delay, the latency introduced by the transmission of the \gls{urllc} preamble and payload are respectively $M_\text{p} T_\text{m}$ and $ M_\text{u} T_\text{m}$. Moreover, we let $D_\text{proc} > 0$ be a constant that accounts for the time needed to process the preamble at the \gls{ris} and detect the \gls{urllc} traffic.
Also, recall that the delay for the \gls{ris} to switch to the \gls{urllc}-oriented configuration is $M_\text{s} T_\text{m}$. Then, the \gls{urllc} latency is:
\begin{equation}
    D = M_\text{p} T_\text{m} + M_\text{u} T_\text{m} + D_\text{proc} + M_\text{s} T_\text{m}.
\end{equation}
If the processing delay is negligible compared to the mini-slot duration, a reasonable approximation for the \gls{urllc} latency is $D \approx (M_\text{p} + M_\text{s} + M_\text{u}) T_\text{m}$.
As a comparison, in the \gls{5gnr} preemptive puncturing, the latency increases in the order of the slot duration due to the \gls{bs} coordination~\cite{3gpp38300}. Due to space limitations, we leave a detailed comparison for future work.

\section{Numerical Results}\label{sec:numerical-results}

Now, we present numerical simulations to discuss the performance of the proposed scheme for multiplexing \gls{embb} and \gls{urllc} services. In the simulations, $\mathbf{h}_\text{b}$ and $\mathbf{h}_\iota$ follow the line-of-sight channel model in \cite[eqs. (6) and (7)]{Albanese2022}, with $\lambda = 0.1$ m, $\beta = 3.67$, $\gamma_0 = 1$, $d_0 = 1$ m, and $\sigma^2 = - 90$ dBm.\footnote{The operations of the proposed algorithms and benchmarks are independent of the channel model, applying to other, more general, models.}
Moreover, we consider $M_\text{p} = M_\text{s} = 1$ mini-slot and $M_\text{u} = 2$ mini-slots. The \gls{ris} is a square surface placed at the origin of the coordinate system
in the $yz$-plane, pointing towards the direction of the $x$-axis. The \gls{bs} is at $\mathbf{q}_\text{BS} = [\varrho_\text{f}/\sqrt{2} ~ \varrho_\text{f}/\sqrt{2} ~ 0]^\transp$, where $\varrho_\text{f} = \lambda (\sqrt{N} - 1)^2 /2$ is the far-field distance of the \gls{ris}.
The region occupied by the \glspl{ue} is a volume defined in spherical coordinates by the set $\{ (\varrho, \vartheta, \varphi) \mid \varrho_\text{f} \leq \varrho \leq 100 \text{ m}, \vartheta_\text{min} \leq \vartheta \leq \vartheta_\text{max}, \frac{3\pi}{2} \leq \varphi \leq 2\pi \}$, where the tuple $(\varrho, \vartheta, \varphi)$ denotes, respectively, the radial distance, the polar, and azimuthal angles. Specifically, the angles $\vartheta_\text{min}, \vartheta_\text{max}$ are such that the $z$-coordinates of the \glspl{ue} lie within $[-3, 3]$~m. The \glspl{ue}' positions, drawn for each $10^7$ realization, are uniformly distributed over this region.

\vspace{1mm}
\noindent\textbf{Benchmarks:} To compare the proposed algorithms for computing the \gls{urllc}-oriented configuration, we define the following benchmarks:
\textbf{a)}~\textit{Random:} The phase shifts $\{ \theta_n^\text{u} \}_{n = 1}^N$ are drawn from a uniform distribution over the interval $[0, 2\pi)$.
\textbf{b)}~\textit{Missed \gls{urllc} preamble:} \gls{embb}-oriented configuration, representing the case where the \gls{ris} always fails to detect the start of the \gls{urllc} traffic ($\epsilon_\text{m} = 1$).\footnote{The benchmarks \textbf{a)} and \textbf{b)} are respectively equivalent to the configurations to assist the \gls{urllc} and \gls{embb} \glspl{ue} in \cite{Souto2021}.}
{\textbf{c)}~\textit{Preemptive puncturing:} Following the procedure described in~\cite{3gpp38300}, the \gls{embb} \gls{tti} is interrupted so as the \gls{urllc} \gls{ue} has an interference-free \gls{tti}. Still, since there is no \gls{urllc} \gls{csi} knowledge at the \gls{bs}, the \gls{ris} keeps the \gls{embb}-oriented configuration.}
{\textbf{d)}}~\textit{Maximize URLLC SNR:} Ideally, assuming that \gls{urllc} \gls{csi} is available at the \gls{bs} prior to its \gls{tti}, $\boldsymbol{\psi}^\text{u}$ is set to perform coherent passive beamforming like in~\eqref{eq:ris-configuration-embb}, but for the \gls{urllc} \gls{ue}.

\begin{figure*}[h]
    \subfloat[]{%
        \includegraphics[width=.25\columnwidth]{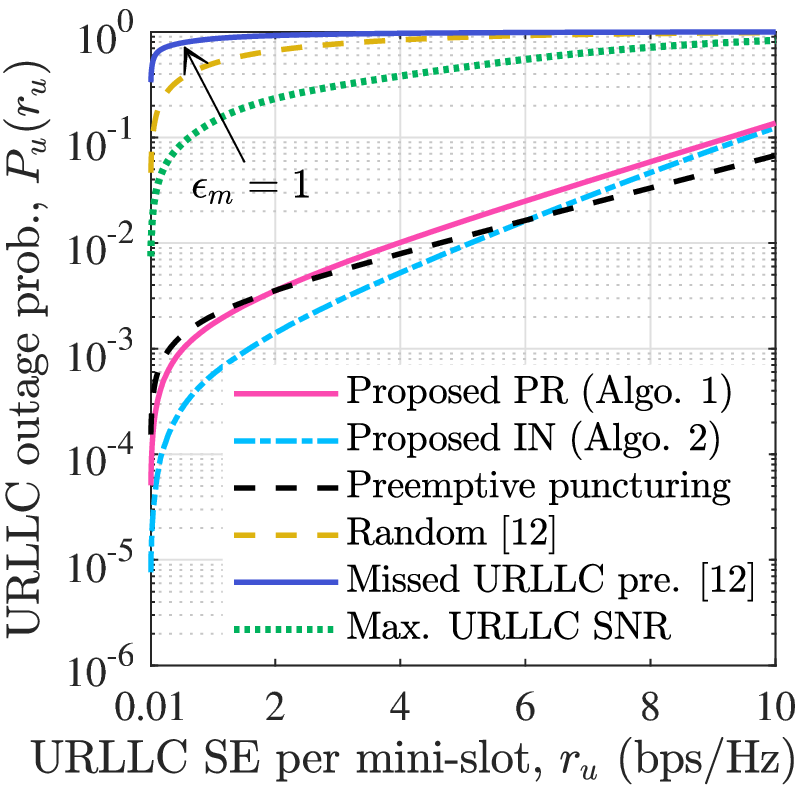}%
        \label{fig:numerical-results-urllc-se}%
    }\hfill%
    \subfloat[$p_\text{u} = 23$ dBm]{%
        \includegraphics[width=.1875\columnwidth]{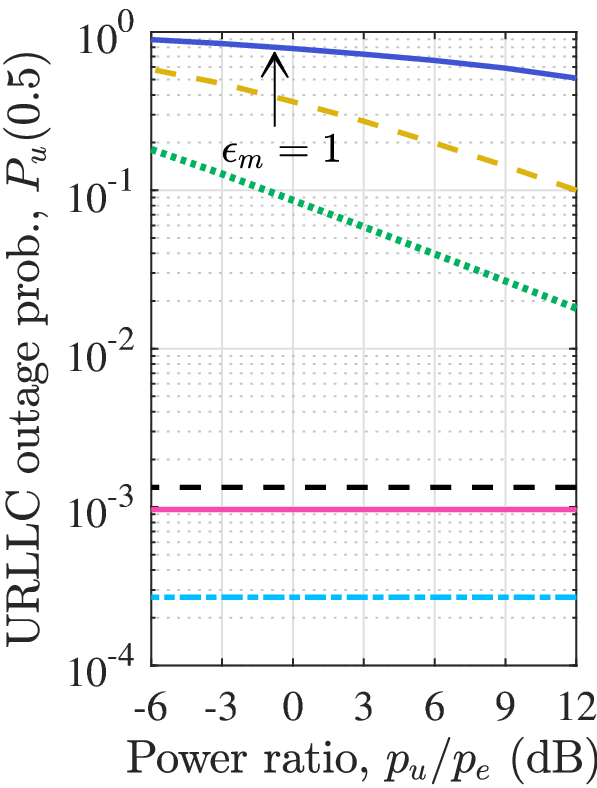}%
        \label{fig:numerical-results-power-ratio}%
    }\hfill%
    \subfloat[]{%
        \includegraphics[width=.1875\columnwidth]{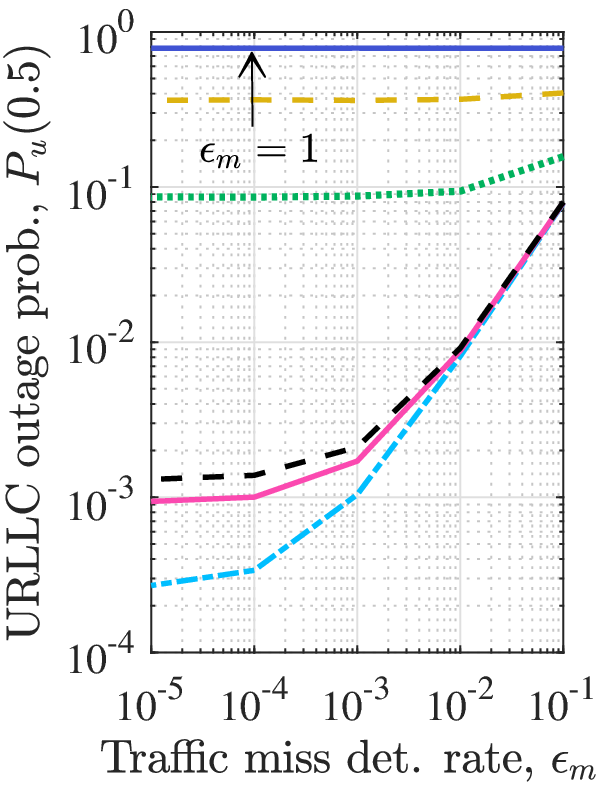}%
        \label{fig:numerical-results-traffic-miss-detection-rate}%
    }\hfill%
    \subfloat[$\varrho_\text{f} = 18$ m]{%
        \includegraphics[width=.1875\columnwidth]{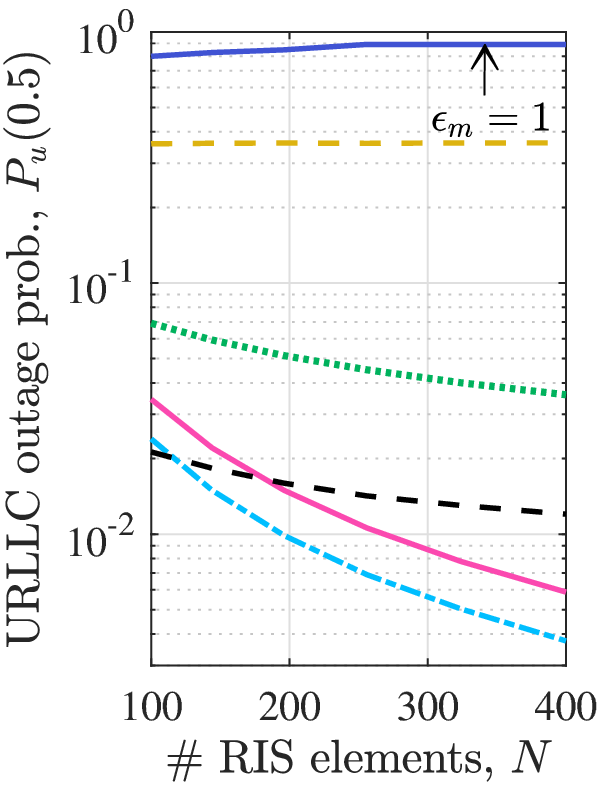}%
        \label{fig:numerical-results-ris-elements}%
    }\hfill%
    \subfloat[$\varrho = 100$ m]{%
        \includegraphics[width=.1875\columnwidth]{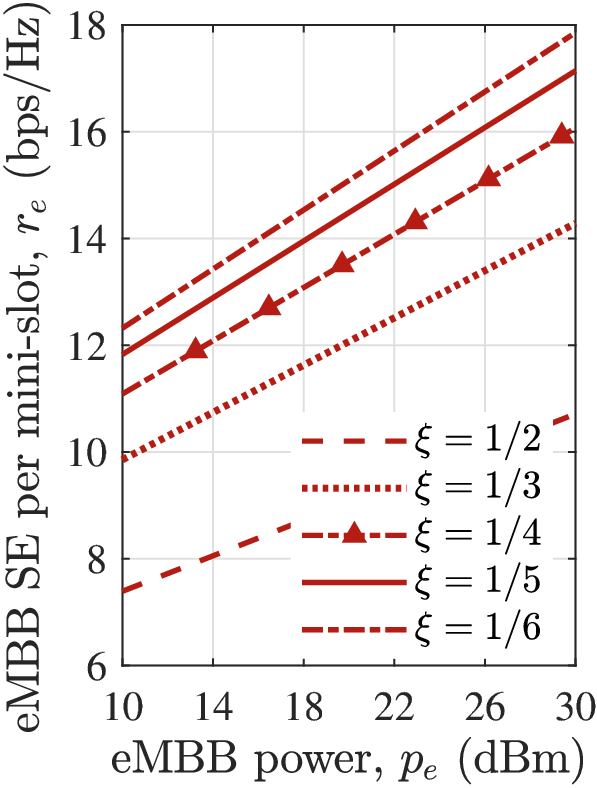}%
        \label{fig:numerical-results-embb-se}%
    }%
    \caption{\gls{urllc} outage probability and \gls{embb} \gls{se} at the cell edge. When not otherwise specified, the transmit powers are $p_\text{e} = p_\text{u} = 23$~dBm, the \gls{ris} has $N = 100$ elements, the \gls{urllc} traffic miss detection rate is $\epsilon_\text{m} = 0$, and the \gls{ris} far-field distance is $\varrho_\text{f} = 4$~m. Also, in (e), $\xi$ is the fraction of \gls{embb} mini-slots unaffected by the \gls{urllc} \gls{tti} and the \gls{ris} switching, as defined in~\eqref{eq:mutual-information-embb}.}
    \label{fig:numerical-results}
\end{figure*}

Figs.~\ref{fig:numerical-results-urllc-se} and~\ref{fig:numerical-results-power-ratio} present the \gls{urllc} outage probability as a function of the \gls{se} and the transmit power ratio, respectively. These results reveal the gains of the proposed \gls{pr} and IN algorithms over the benchmarks, even relying only on \gls{embb} \gls{csi}. Notice that the \gls{pr} and IN performance is comparable to the preemptive puncturing, where there is no \gls{embb} interference due to the \gls{bs} scheduling.
Remarkably, the IN algorithm outperforms the preemptive puncturing performance by 4.9~times in Fig.~\ref{fig:numerical-results-power-ratio}.
Compared to the random and the \gls{pr} configurations, \gls{in} reduces the outage probability by up to 3~orders of magnitude and 3.5~times, respectively. However, the \gls{pr} algorithm yields a better trade-off due to its simpler implementation.

Fig.~\ref{fig:numerical-results-traffic-miss-detection-rate} depicts the \gls{urllc} outage probability as a function of the \gls{urllc} traffic miss detection rate. {Only for preemptive puncturing, $\epsilon_\text{m}$ is considered to be the failure rate for the \gls{urllc} scheduling request procedure.} Notice that, for the proposed scheme, the traffic detection scheme is not considered a major limiting factor for the performance, mainly because there are simple architectures for it that yield extremely low detection error rates. For example, a scheme in~\cite{Jiang2022} achieves a detection error rate of up to $10^{-7}$ at an \gls{snr} of 4~dB with a preamble comprised by a single \gls{ofdm} symbol.

Fig.~\ref{fig:numerical-results-ris-elements} depicts the \gls{urllc} outage probability as a function of the number of \gls{ris} elements. It is worth mentioning that, as the outage probability improves with a bigger surface, the overhead for estimating the \gls{embb} \gls{csi} increases proportionally to $N$. Hence, when selecting the number of \gls{ris} elements, a trade-off exists between the \gls{urllc} outage probability and the overhead in the \gls{csi} estimation phase.

Fig.~\ref{fig:numerical-results-embb-se} depicts the {\gls{se} of the \gls{embb} \gls{ue} when placed} at the cell edge ($\varrho = 100$~m) as a function of the transmit power. Notice that, as the \gls{urllc} packet size is relatively small compared to the \gls{embb}, the mini-slots erased by the \gls{urllc} traffic have minimal impact on the \gls{embb} \gls{tti}, resulting in high \gls{embb} \gls{se}.

\section{Conclusions}

In this letter, we have proposed an \gls{ris}-assisted \gls{ul} multiplexing scheme to support the \gls{embb}\slash\gls{urllc} coexistence. The scheme relies on two \gls{ris} configurations computed from \gls{embb} \gls{csi}. The first configuration is a coherent passive beamformer to maximize the \gls{embb} \gls{snr}. The second is computed by the proposed \gls{pr} and \gls{in} algorithms and mitigates the \gls{embb} interference in the \gls{urllc} traffic, temporarily silencing the \gls{embb} traffic. Numerical results have demonstrated that the proposed scheme enables a balanced coexistence of the services in the absence of \gls{urllc} \gls{csi}.

\bibliographystyle{IEEEtran}
\bibliography{references}

\end{document}